\documentclass[twocolumn,showpacs,floats,floatfix,superscriptaddress,aps,pra]{revtex4-1}
\usepackage{amsfonts}
\usepackage{amssymb}
\usepackage{amsmath}
\usepackage{graphicx}
\usepackage{bm}
\usepackage{color}
\usepackage{epsfig}
\usepackage{calc}
\usepackage{ifthen}
\usepackage{subfigure}
\usepackage{graphicx}
\usepackage{epstopdf}
\usepackage{epsfig}

\begin{document}

\title{Adiabatic following of terahertz surface plasmon-polaritons based on tri-layered corrugated thin film coupler}
\date{\today }

\begin{abstract}
In this paper, we utilize coupled mode theory (CMT) to model the coupling of surface plasmon-polaritons (SPPs) between tri-layered corrugated thin films (CTF) structure coupler in the terahertz region. 
Employing the stimulated raman adiabatic passage (STIRAP) quantum control technique, we propose a novel directional coupler based on SPPs evolution in tri-layered CTF in some curved configuration.
Our calculated results show that the SPPs can be completely transferred from the input to the output CTF waveguides, and even we consider SPPs propagation loss, the transfer rate is still above $70 \%$. The performance of our coupler is also robust that it is not sensitive to the geometry of device and wavelength of SPPs. As a result, our device can tolerate defect induced by fabrication and manipulate THz wave at broadband.

\end{abstract}

\pacs{}
\author{Wei Huang}
\affiliation{Guangxi Key Laboratory of Optoelectronic Information Processing, Guilin University of Electronic Technology, Guilin 541004, China}

\author{Shan Yin}
\email{syin@guet.edu.cn}
\affiliation{Guangxi Key Laboratory of Optoelectronic Information Processing, Guilin University of Electronic Technology, Guilin 541004, China}

\author{Wentao Zhang}
\email{zhangwentao@guet.edu.cn}
\affiliation{Guangxi Key Laboratory of Optoelectronic Information Processing, Guilin University of Electronic Technology, Guilin 541004, China}

\author{Kaili Wang}
\affiliation{Guangxi Key Laboratory of Optoelectronic Information Processing, Guilin University of Electronic Technology, Guilin 541004, China}

\author{Yuting Zhang}
\affiliation{Guangxi Key Laboratory of Optoelectronic Information Processing, Guilin University of Electronic Technology, Guilin 541004, China}

\author{Jiaguang Han}
\affiliation{Center for Terahertz Waves and College of Precision Instrument and Optoelectronics Engineering, Tianjin University, Tianjin 3000072, China}

\maketitle



\section{Introduction}


Terahertz (THz) radiation has drawn enormous attentions these years. Since many material responses are located at THz frequency, THz technologies can obtain unique spectral characteristics and abundant information about matters, which is widely used in spectroscopy \cite{Lee2009} and imaging \cite{Chan2007}. Naturally, the THz applications in information processing and transmission \cite{Ozbay2006,Lee2010} are vital. On the other hand, with the rapid development of the network and popularization of portable terminals, the miniaturization of the integrated devices is an irresistible trend. THz technologies are promising to accelerate the next generation of communications \cite{Naeem2018, Withayachumnankul2018, Yu2016} due to the capabilities of high capacity and micro-size \cite{Koenig2013,Ostmann2011}. To realize further integration, how to manipulate the electromagnetic (EM) waves in subwavelength scale is a key issue. Surface plasmon polaritons (SPPs) are the EM waves propagating along metal-dielectric interfaces with exponential decay in the direction perpendicular to the interfaces. The recent emerged SPPs-based elements, such as antennas \cite{Schnell2009,Maguid2016}, waveguides \cite{Sorger2011,Ebbesen2008} and logic circuitry \cite{Ebbesen2008,Cohen2013}, demonstrated their potential application on the microscale and nanoscale chips since the wavelength of SPPs can be scaled down below diffraction limit \cite{Maier2007,Gramotnev2010,Kawata2009}. 

At terahertz regime, SPPs-based waveguides \cite{Zhang2017}, couplers \cite{Ma2017} and coders \cite{Yin2018} have been investigated recently, which will make great contributions to the THz applications. Due to these advantages of SPPs at terahertz regime, completely transfer energies and information of THz SPPs is significant to implement compact device in THz regime. Two recent researches studying on the coupling of THz SPPs waveguides \cite{Liu2014,Zhang2018} involved in coupled mode theory (CMT), which is a widely used theory in describing coupling between two optical waveguides, through the overlap of their evanescent electromagnetic fields \cite{Yariv1973,Huang2014}. Base on this concept, if two thin films are close enough, the two evanescent fields of SPPs in each thin film have overlapping and SPPs can transfer from one thin film to another \cite{Liu2014,Zhang2018}. In our paper, we employ and derive the CMT to describe the SPPs coupling between courrgated thin films structure. 

However, the present stuctures of two parallel THz SPPs waveguides (e.g. ref. \cite{Liu2014,Zhang2018}) require rigorous fabrication precision and only operate at specific excited frequency of THz waves, otherwise, the fidelity of device will drop rapidly. Most recently, to overcome this shortcoming, a remarkable paper applied coherent quantum control (stimulated raman adiabatic passage, short for STIRAP) into transferring the SPPs on the graphene sheets \cite{Huang20181}. STIRAP is the well-known three-level coherent quantum control, which provides completely transfer population from first state to third state, without any population remaining In intermediate state \cite{Vitanov20011,Vitanov20012,Vitanov2017}. Furthermore, it is shown that STIRAP is exceedingly robust against controlling parameters under perturbations. The SITRAP has already widely used in various domains, such as atomic molecular and optical physics \cite{Yale2016,Huang2017}, waveguide coupler \cite{Mrejen2015,Longhi2007}, graphene electronic and optical effect \cite{Huang20181,Huang20182}. In this paper, we firstly introduce the STIRAP technique into the SPPs waveguide coupler at terahertz regime, to achieve very robust device against varying frequency of input THz waves and disturbances on the geometry parameters. We propose the tri-layered corrugated thin film coupler structure with some curved configuration and we substantiate that the performance of our coupler is also robust to the geometry of device and wavelength of SPPs. As a result, our device can tolerate defect induced by fabrication and manipulate THz wave at broadband, which is meaningful in developing THz functional devices.

\section{Model}

We first consider terahertz radiation to excited surface plasmon-polaritons on the surface of the courrgated thin films structure. Assuming a slab courrgated thin films locates at $z=0$ at $x z$ plane, we illuminate the terahertz waves on the surface of the thin film to excited SPPs propagating along $x$ direction. In order to SPPs extend the propagation distance, it is remarkable to utilize courrgated structure cutting on the thin film \cite{Zhang2017,Liu2014,Zhang2018}, as shown in Fig. 1, with cutting depth $h$, width $a$, period $d$ and thickness of thin film $t$. If we contemplate the mode profile of SPPs, electric field of SPPs has exponentially decay along with $y$ and $z$ directions outside the SPPs waveguide, as the evanescent field of electric field. 

In this paper, we only study the coupling mechanism along $z$ direction. Therefore, SPPs' electric fields of $x$ direction (SPPs propagation) and $z$ direction (SPPs' evanescent field) are observed and we ignore the impacts of $y$ direction. Assume that we place two corrugated thin films at $z=g/2$ and $-g/2$ and these two films are parallel to $x y$ plane, where $g$ is the gap distance between two parallel thin films (see Fig. 1). The TM polarized SPPs modes are excited on one corrugated thin films. The electric fields can be described by $E_1 = (E_{1x}, 0, E_{1z}) e^{iqx} e^{-k_m |z-g/2|}$ and $E_2 = (E_{2x}, 0, E_{2z}) e^{iqx} e^{-k_m |z+g/2|}$. Here $k_m$ is the decay rate of evanescent field in the surrounding dielectric mediums, given by $k_m = \sqrt{ (\omega^2 \epsilon_m - q^2) /c^2}$ \cite{Saleh1991}. $\epsilon_m$ is the permittivity of medium material (we use silicon as surrounding mediums) and $\omega$ is the frequency of incident light in air. In addition, $q$ is the propagation constant of SPPs and it well depends on the geometry structure and frequency of incident terahertz light $\omega$ \cite{Zhang2017,Ma2017,Maier2006}. We can numerically solve it by dispersion equation, given by $q=\frac{\omega}{c}\sqrt{1+\frac{a^2}{d^2} \tan^2 \frac{\omega h}{c}}$ \cite{Maier2006}. 

 \begin{figure}[hbtp]
\centering
\includegraphics[width=0.5\textwidth]{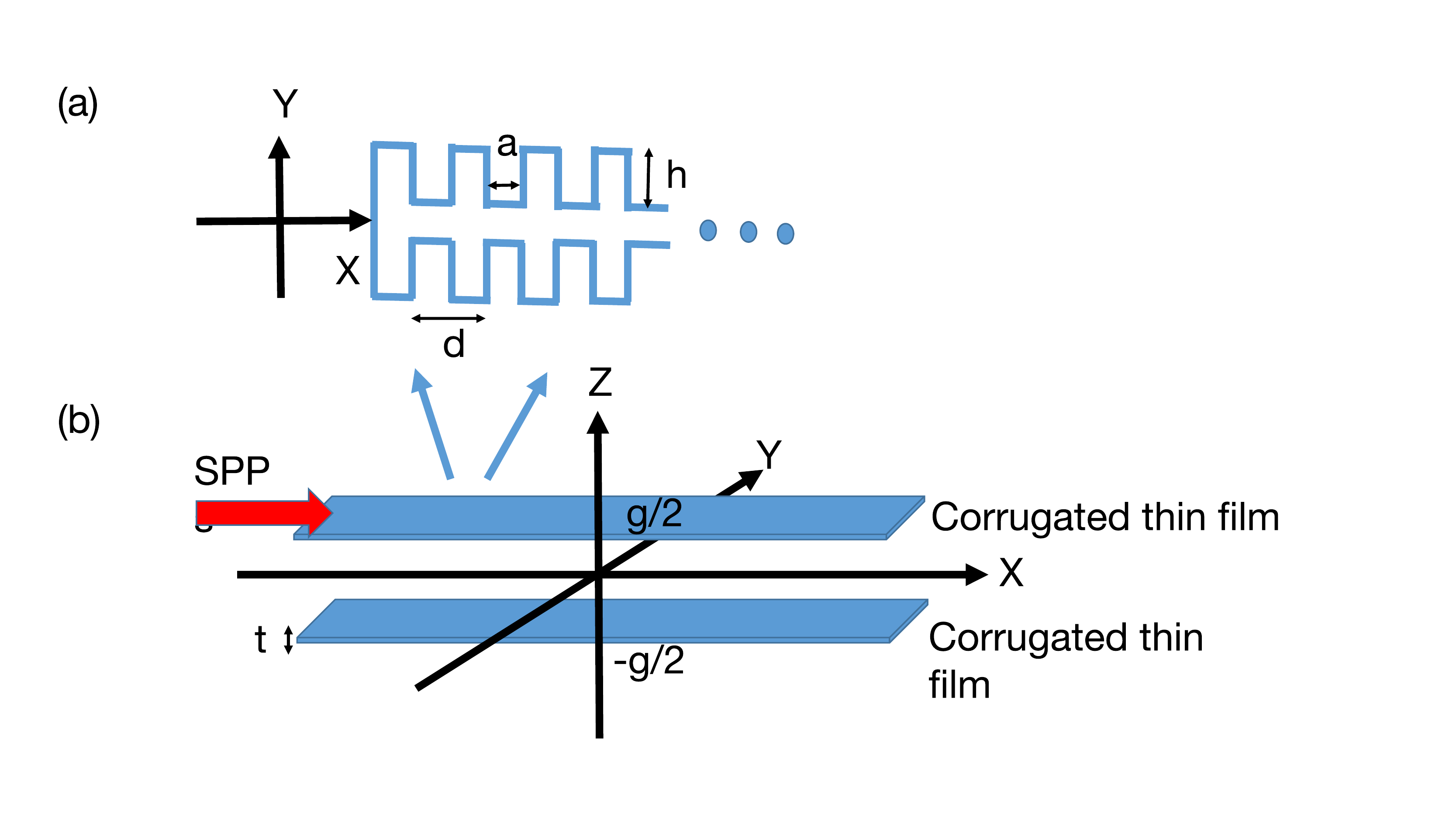}
\caption{(a) the enlarge view of the corrugated structure of thin film. (b) two parallel corrugated thin film place at $z=g/2$ and $z=-g/2$, where $g$ is the gap distance between two thin films. }
\end{figure}

In our parallel coupling model, we take the notations $\Psi_1(x,z)$ ($\Psi_2(x,z)$) as the electric field of SPPs on the first (second) thin film, written as

\begin{equation}
\begin{aligned}
\Psi_1(x,z)= a_{1}(x) u_{1}(z) \exp(-i q x), \\
\Psi_2(x,z)= a_{2}(x) u_{2}(z) \exp(-i q x),
\end{aligned}
\end{equation}
where $a_{1}(x)$ and $a_{2}(x)$ are the amplitudes of the modes with respect to SPPs on two thin film. Due to the extremely thickness of film ($t = 100$ nm) comparing to other geometry parameters, we can obtain the mode profiles of SPPs as $u_{1} = E_{1z} exp(- k_m |z-g/2|)$ and $u_{2} = E_{2z} exp(- k_m |z+g/2|)$. The electric filed have to normalize by the normalization factor as $N_{1,2} = \sqrt{ \int^{+\infty}_{-\infty} |u_{1,2}(z)|^2 dz}$ for the thin film 1 and 2, respectively. We take the notation as $\psi_{1}$=$u_{1}(z) \exp(-i q x)$ and $\psi_{2}$=$u_{2}(z) \exp(-i q x)$, where $\psi_{1}$ and $\psi_{2}$ must be satisfied Helmholtz equations in the $x$ direction.

Based on the CMT model, we can manipulate the Helmholtz equations with the source terms to obtain
\begin{equation}
\begin{aligned}
\dfrac{\partial^2}{\partial x^2} \Psi_{1}(x,z) + q^2 \Psi_{1}(x,z) = -(k_{2}^2 - k_0^2)\Psi_{2}(x,z), \\
\dfrac{\partial^2}{\partial x^2} \Psi_{2}(x,z) + q^2 \Psi_{2}(x,z) = -(k_{1}^2 - k_0^2)\Psi_{1}(x,z),
\end{aligned}
\end{equation}
where $k_0=\sqrt{(\omega^2 \epsilon_{g} - q^2 )/ c^2}$ with thin film (gold) permittivity $\epsilon_{g}$. These equations are consistent with the conventional optical waveguide coupled equations \cite{Saleh1991}.

By substituting the wave functions of the SPPs on two thin films into the given Helmholtz equations, we simplify the formation by using the slowly varying envelope approximation \cite{Saleh1991}, namely $\frac{d^2 a_1}{dx^2} \ll \frac{d a_1}{dx}$ and $\frac{d^2 a_2}{dx^2} \ll \frac{d a_2}{dx}$.
Under this approximation, the coupling equations can be rewritten as a Schr\"odinger-like equation of a two-level system, given by
\begin{equation}
i\dfrac{d}{d x}
\begin{bmatrix}
a_{1} \\
a_{2}
\end{bmatrix}
= \begin{bmatrix}
0 & C_{12} \\
C_{21}  & 0
\end{bmatrix} \begin{bmatrix}
a_{1} \\
a_{2}
\end{bmatrix}.
\end{equation}
Here, $C_{12}$ and $C_{21}$ are the coupling coefficients: $C_{12} = \frac{1}{2} \frac{k_{2}^2 - k_0^2}{q}  \int^{+\infty}_{-\infty} u_{1}(z) u_{2}(z) dx$ and $C_{21} = \frac{1}{2} \frac{k_{1}^2 - k_0^2}{q}  \int^{+\infty}_{-\infty} u_{1}(z) u_{2}(z) dx$.

 \begin{figure}[hbtp]
\centering
\includegraphics[width=0.5\textwidth]{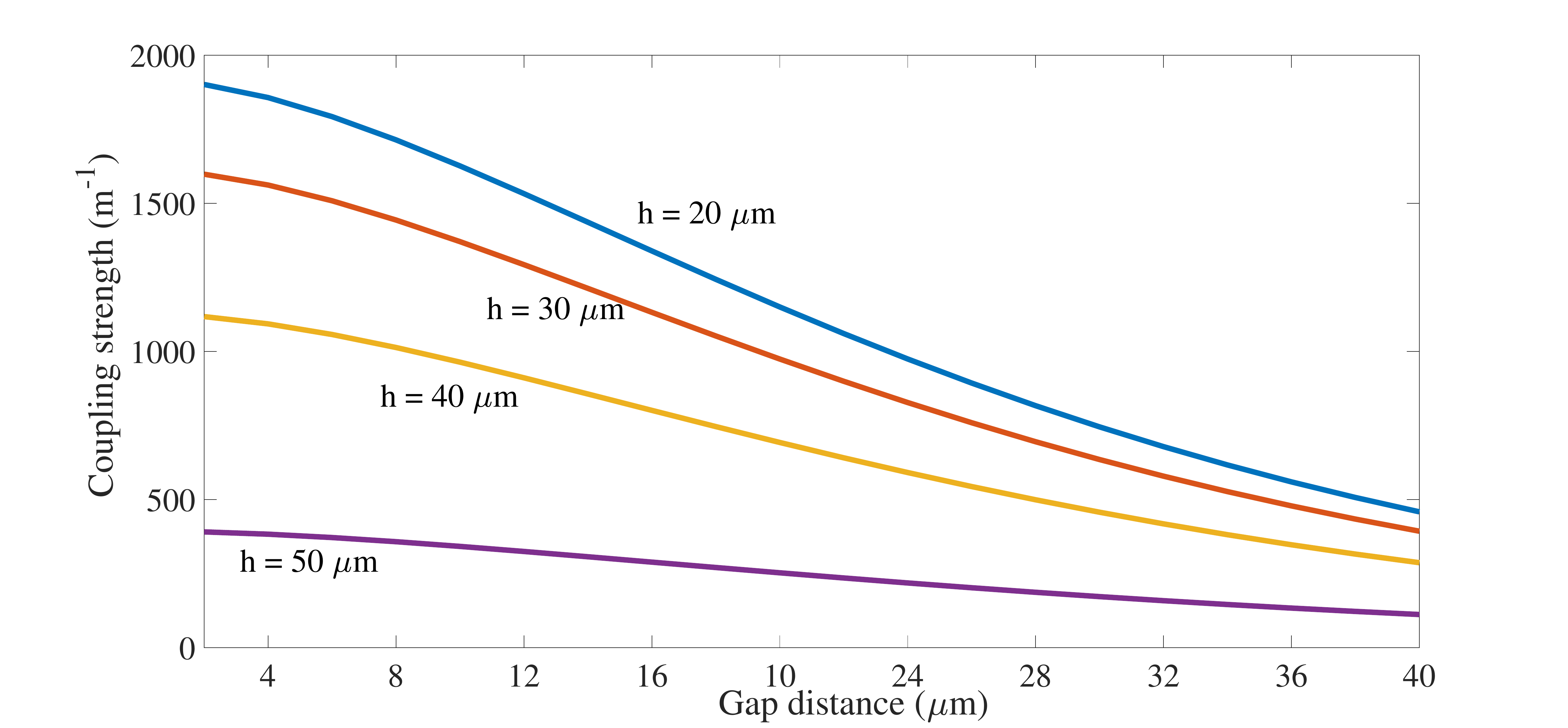}
\includegraphics[width=0.5\textwidth]{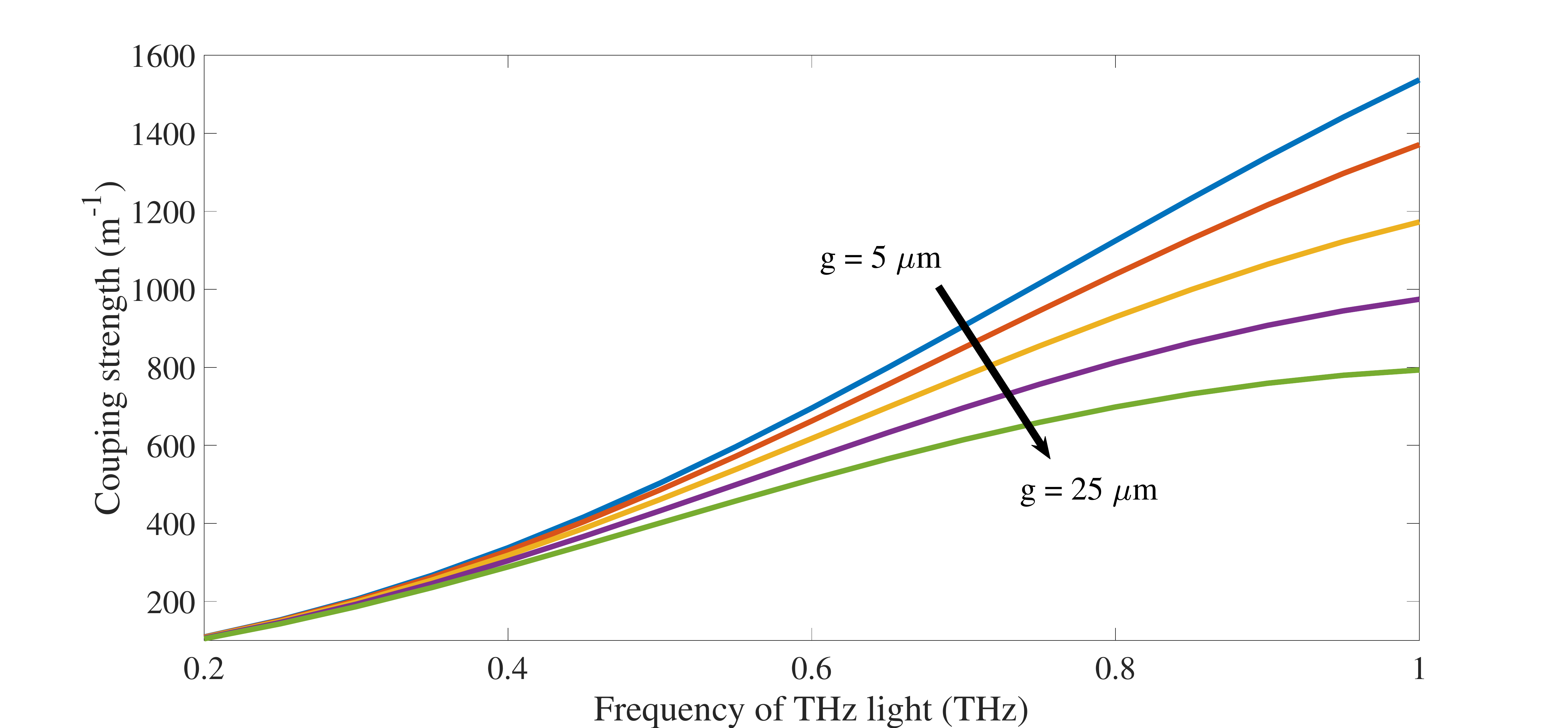}
\caption{(a) The coupling strength against varying gap distance and height of corrugated structure. (b) The function of coupling strength with changing frequency of input Terahertz radiation and gap distance. }
\end{figure}

As an example, it is straight forward to obtain the coupling strength $C_{12} = C_{21} = 1255$ $m^{-1}$, with setting up $a = 40$ $\mu m$, $d = 50$ $\mu m$, $g = 4$ $\mu m$ and input frequency of terahertz waves at 1 THz. Beneficial to illustrate trends of the coupling strength against to different parameters, we demonstrate coupling strength against varying gap distance and height of corrugated structure (see Fig 2a). Furthermore, Fig. 2b shows the coupling strength as the function of changing frequency of input terahertz radiation and gap distance. From results of Fig. 2a, the coupling strength will exponentially decrease either by increasing gap distance or by increasing the depth courrgated structure $d$. It is remarkable to attain that escalating the frequency of input terahertz radiation will enhance the coupling strength from 0.2 THz to 1 THz, shown in Fig. 2b. These results are noticeably the same as the trends of the SPPs' coupling strength on two parallel thin film structure, Ref. \cite{Liu2014}. 

In our first example, we only consider SPPs coupling between two corrugated thin films. Based on the coupling equation of bi-layered SPPs coupling (Eq. 3), it is obviously to extend the SPPs coupling between multi-layered thin films by using analogy of multi-level Schr\"odinger equation, written as

\begin{equation}
i\dfrac{d}{d x}
\begin{bmatrix}
a_{1} \\
\vdots \\
a_{n}
\end{bmatrix}
= \begin{bmatrix}
0 & C_{12}(x) & \ddots \\
C_{21}(x) & \ddots & C_{n-1,n}(x) \\
\ddots & C_{n, n-1}(x) & 0 \\
\end{bmatrix} \begin{bmatrix}
a_{1} \\
\vdots \\
a_{n}
\end{bmatrix}.
\end{equation}
Here, $C_{12}(x)$ ($C_{21}(x)$) is the coupling SPPs between first and second layer courrgated thin film and $C_{n-1,n}(x)$ ($C_{n,n-1}(x)$) is the coupling SPPs between $(n-1)^{th}$ and $n^{th}$ layer thin film. Similarly, $a_1$ and $a_n$ are the SPPs amplitudes of first and $n^{th}$ courrgated thin film.

\section{Adiabatic following SPPs on tri-layered thin film coupling}

In our novel design of SPPs' adiabatic following, we employ tri-layered SPPs on the corrugated thin film coupler. The geometry scheme of our structure is shown in Fig 3. We have tri-layered of corrugated thin film, which are placed at $x z$ plane to obtain the $z$ directional coupling and $x$ direction SPPs propagation. The middle layer is flat courrgated thin film, located at $x = 0$. The input (first) layer of courrgated thin film has slight curve with radius $R$ and its center is above $x$ axis. The minimum distance between input and middle layer takes the notation as $d_{min}$. After that, we  slightly bend the output (third) layer courrgated thin film with radius $R$, whereas its center is below $x$ axis. $d_{min}$ is also the minimum distance between middle and output layer. Notice that the offset between two centers of circles (input and output layers) in the $x$ axis is $\delta$ and center of output layer is in front of input layer in the $x$ direction. The spatial dependence of the spacing $d_1(x)$ and $d_2(x)$ of the input and output courrgated thin film with respective to the middle layer is given by $d_1(x)=\sqrt{R^2-(x-\delta/2)^2}+(d_{min}+R)$ and $d_2(x)=\sqrt{R^2-(x+\delta/2)^2}-(d_{min}+R)$. Therefore, at the beginning, we excited the SPPs on the input layer of corrugated thin film and the all power of SPPs will completely transfer from input to output layer of corrugated thin film via coupling mechanism by our designing structure. 

 \begin{figure}[hbtp]
\centering
\includegraphics[width=0.5\textwidth]{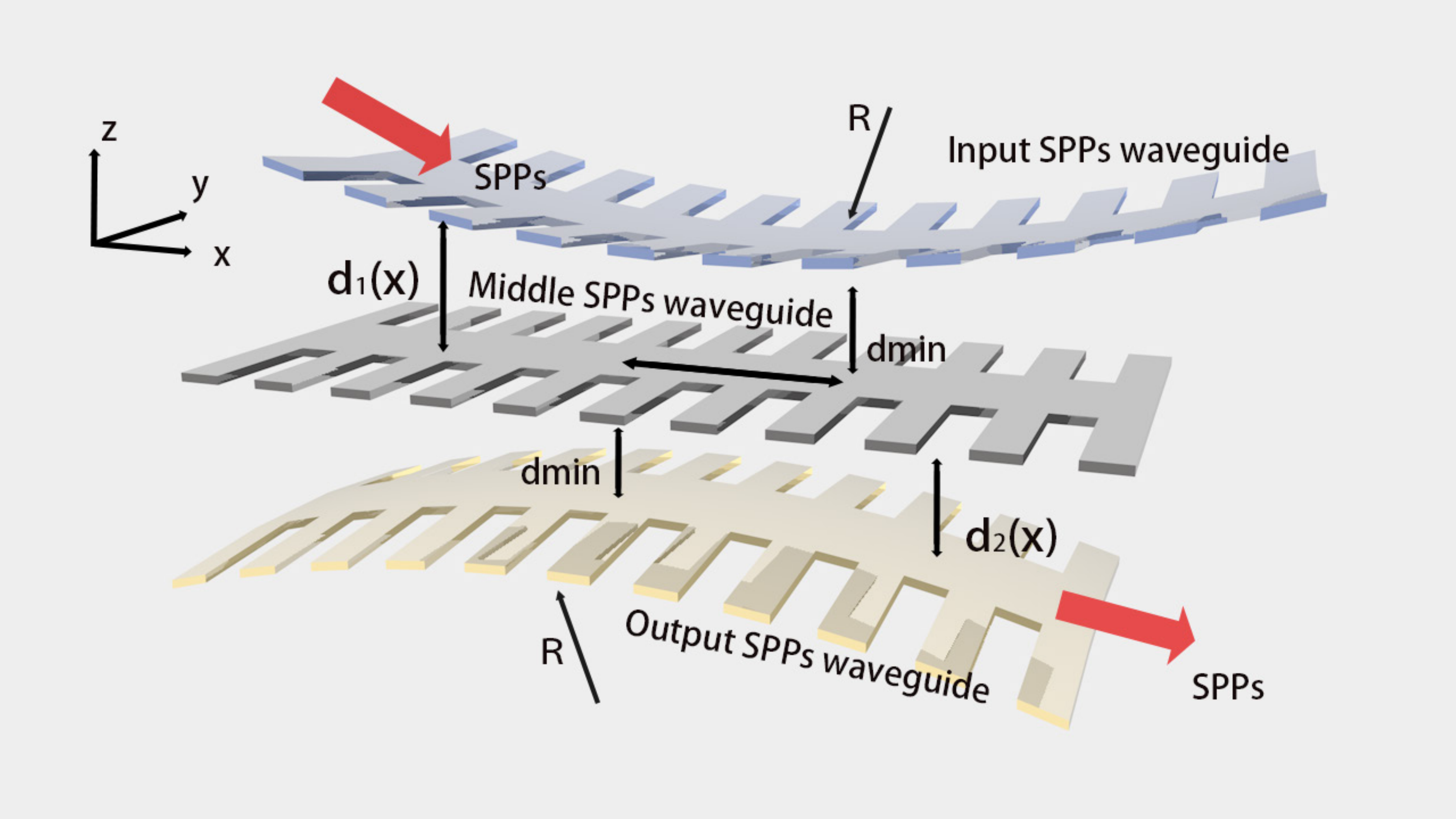}
\caption{The schematic configuration of our designed tri-layered SPPs corrugated thin film coupler based on STIRAP.}
\end{figure}

Thus, based on the multi-layered SPPs coupling (Eq. 4), the coupling equation for tri-layered SPPs coupler is described as,
\begin{equation}
i\dfrac{d}{d x}
\begin{bmatrix}
a_{1} \\
a_{2} \\
a_{3}
\end{bmatrix}
= \begin{bmatrix}
0 & C_{12} & 0 \\
C_{21}  & 0 & C_{23} \\
0 & C_{32} & 0
\end{bmatrix} \begin{bmatrix}
a_{1} \\
a_{2} \\
a_{3}
\end{bmatrix},
\end{equation}
where $a_1$ ($a_2$, $a_3$) is the power amplitude of the first (second, third) SPPs waveguide. $C_{12}$ ($C_{23}$)  is the coupling strength between first and middle layer (middle and third layer), where $ C_{12} = C_{21}$ and $ C_{23} = C_{32}$. 

\begin{figure}[hbtp]
\centering
\includegraphics[width=0.5\textwidth]{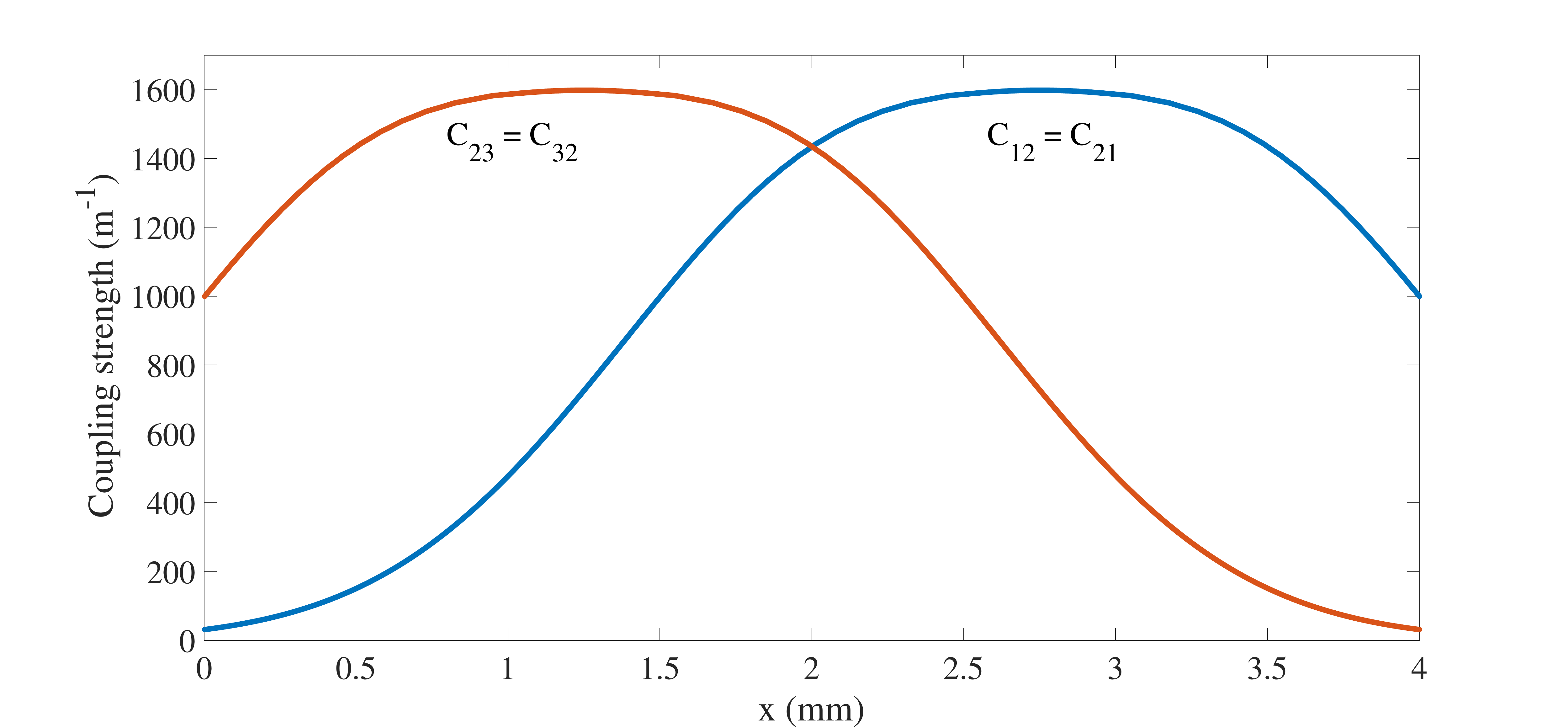}
\includegraphics[width=0.5\textwidth]{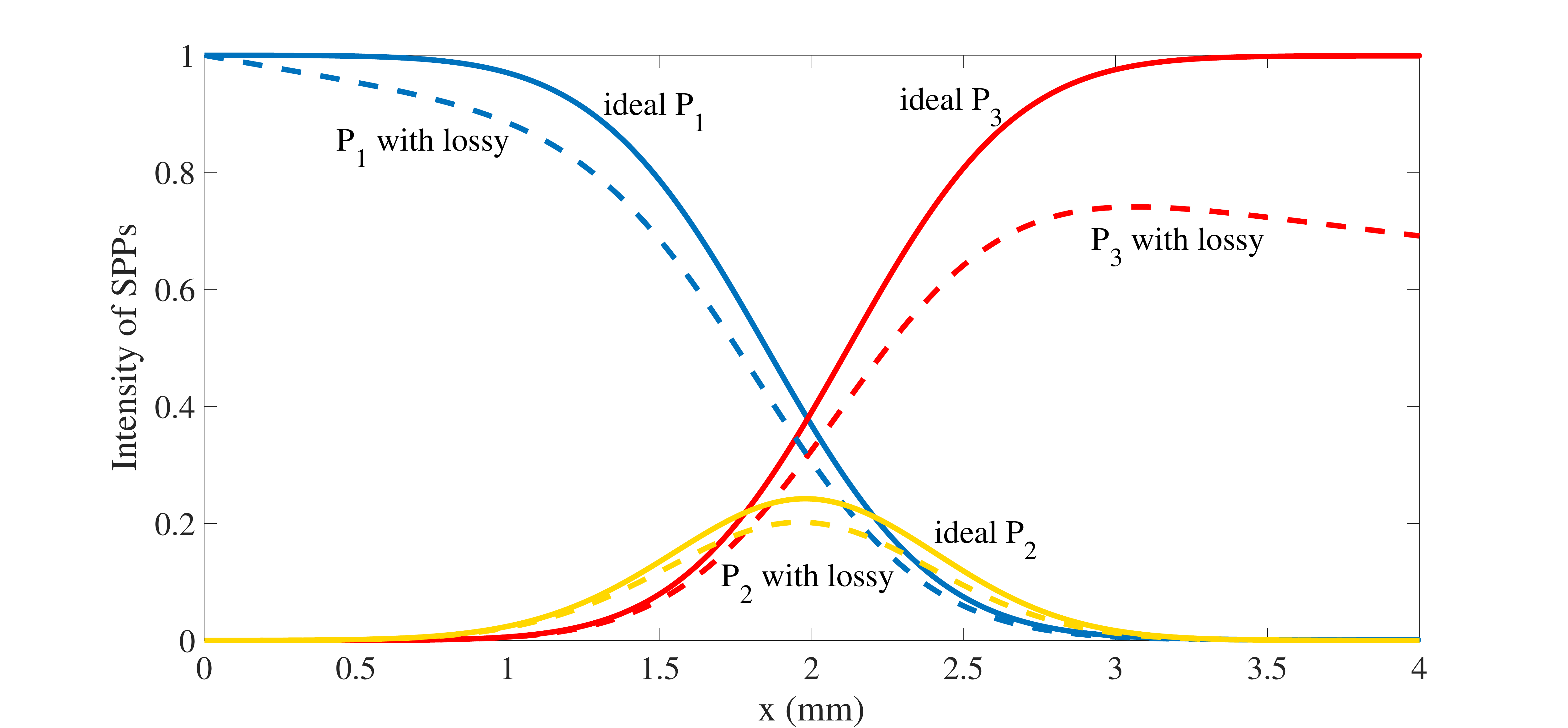}
\includegraphics[width=0.5\textwidth]{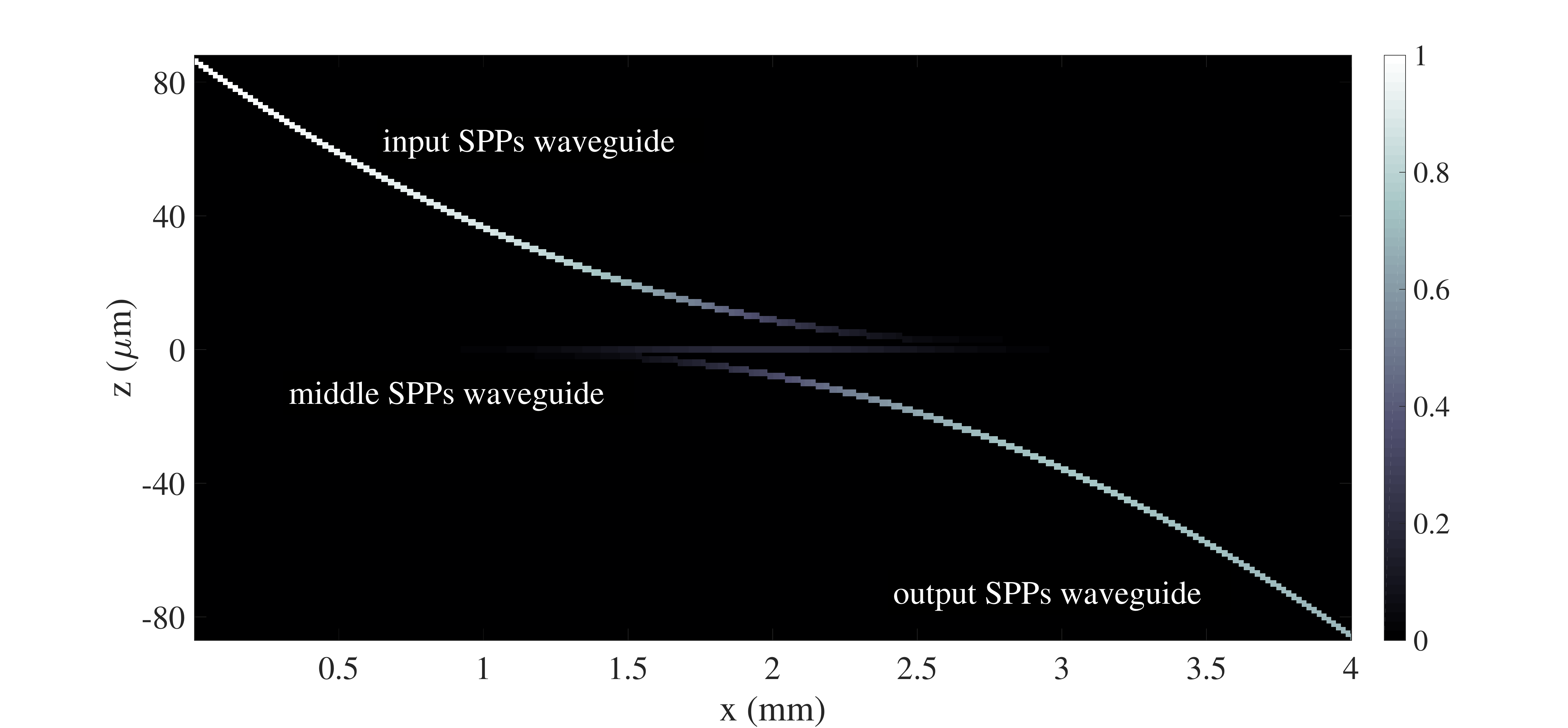}
\caption{(a) The coupling strengths between input and middle $C_{12} = C_{21}$ with blue line (middle and output $C_{23} = C_{32}$ with red line). (b) The intensity of SPPs for first, middle and third corrugated thin film waveguide along with SPPs propagation direction. (c)The visualizing simulation of SPPs propagating complete transfer from first(input) to third(output) corrugated thin film. }
\end{figure}

With the geometry structure of our adiabatic following design (see Fig. 3), we set the geometry parameters as $R = 45$ mm, device length as $L = 4$ mm and distance between two maximum coupling point $\delta = 1.5$ mm. Based on these geometry parameters, we can obtain the coupling strength between input and middle layer $C_{12} = C_{21}$ (middle and output layer $C_{23} = C_{32}$) SPPs waveguide, as shown in Fig. 4a. At the beginning of the transition, the coupling strength of input and middle layer $C_{12} = C_{21}$ is much larger than coupling strength of output and middle layer $C_{23} = C_{32}$. Eventually, the coupling strength $C_{12} = C_{21}$ is much smaller than $C_{23} = C_{32}$. Thus, the whole transition is the quintessential STIRAP transition. The evolution intensities of SPPs within input, middle and output waveguides are shown in Fig. 4b. From the result of Fig. 4b, it is conspicuously to acquire that the intensity of input waveguide ($P_1$) completely transfer to output SPPs waveguide ($P_3$) in the ideal case (without lossy, shown with solid line). However, there is some lossy during SPPs propagation within the corrugated thin film coupler, which is settled as 8 dB/cm in this paper \cite{Zhang2017}. Therefore, the results of SPPs intensity transition with lossy are shown with dashed line in Fig. 4b. We still can achieve efficient transfer of intensity above $70 \%$. The corresponding visualized results with population transfer and geometry structure is demonstrated in Fig. 4c. This intuitionistic outcome illustrates the the SPPs propagation within our adiabatic device, with geometry structure of input, middle, output SPPs waveguides. 

 \begin{figure}[hbtp]
\centering
\includegraphics[width=0.5\textwidth]{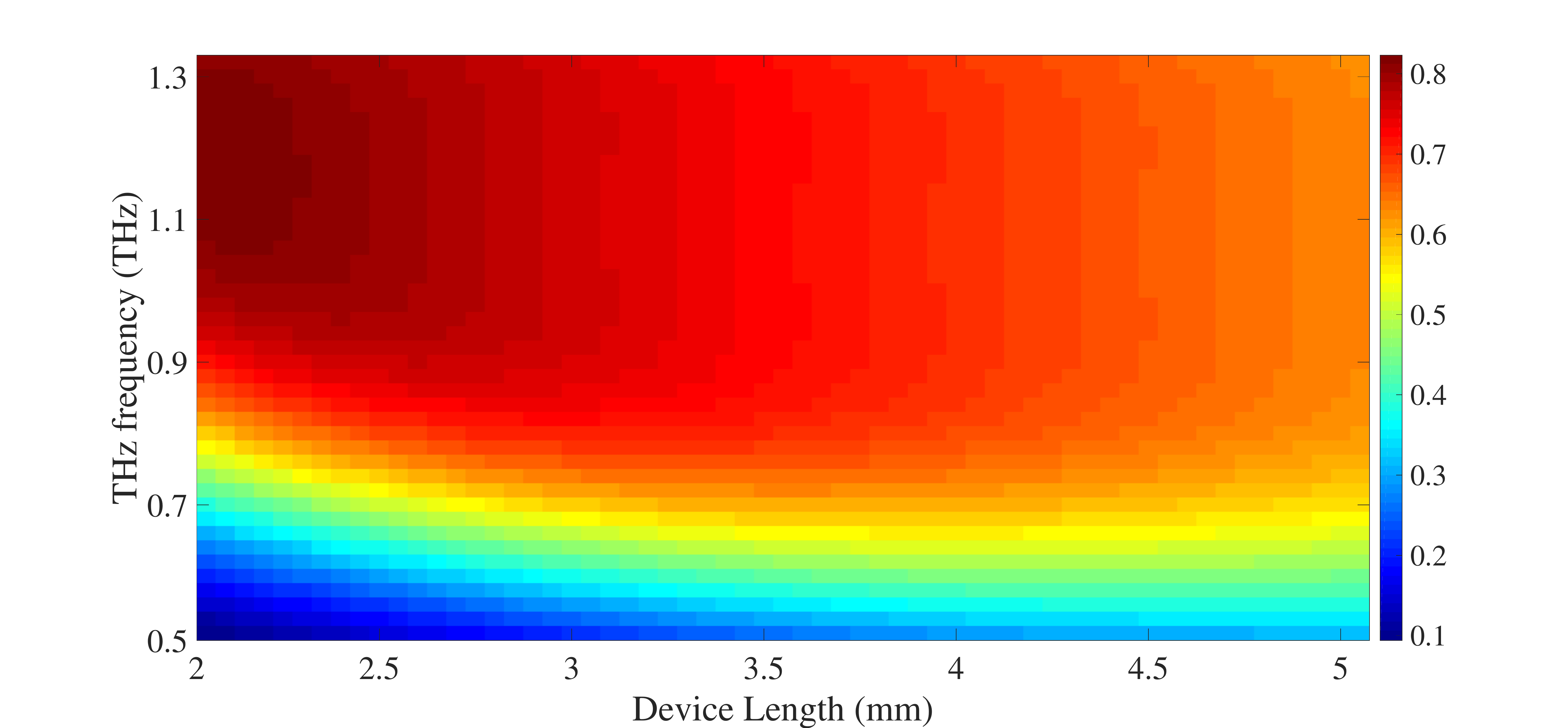}
\includegraphics[width=0.5\textwidth]{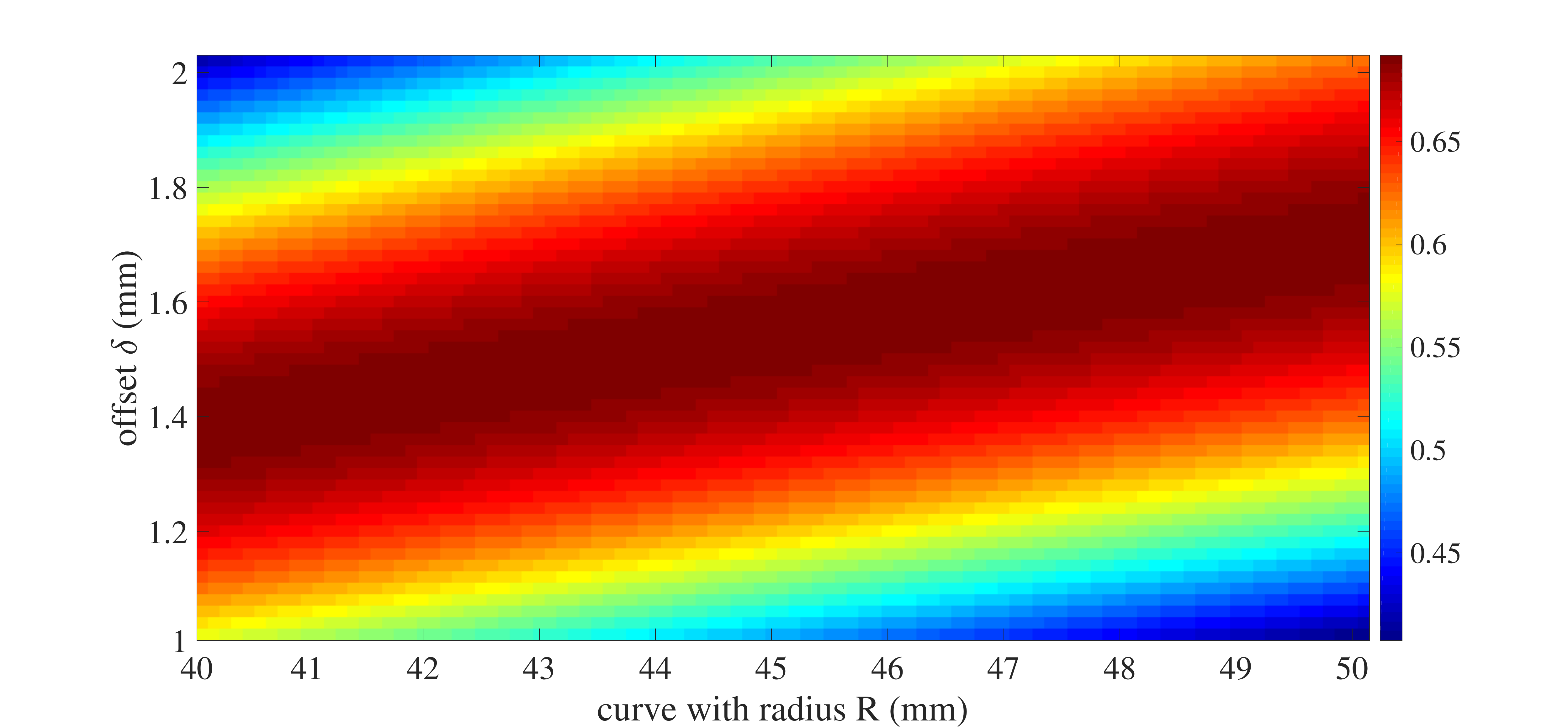}
\caption{(a) The robustness of our adiabatic device, by varying against frequency of input THz light and device length. (b) Varying the offset between two centers of curve $\delta$ (from 1 mm to 2 mm) and curve with radius $R$ (from 40 mm to 50 mm), with setting device length L = 4 mm and exciting by 1 THz  terahertz wave.}
\end{figure}

To authenticate the robustness (by varying against input frequency of THz light and geometry parameters) of our adiabatic following design, we lay out the contour plot of final population at output SPPs waveguide, by scanning the frequency of frequency of input THz light (from 0.5 THz to 1.3 THz) and device length (from 2 mm to 5 mm), shown as Fig. 5a. We conclude that our design can suffer broadband frequency of input THz light (roughly from 0.9 THz to 1.3 THz) and larger perturbation of device length do not deteriorate our performance, which can endure length from 2 mm to 4 mm (energy transfer rate larger than 0.6, even the lossy in consideration). Furthermore, we set the device length L = 4 mm and frequency of THz wave with 1 THz. We plot the final intensity of SPPs in output SPPs waveguide with varying the offset between two centers of curve $\delta$ (from 1 mm to 2 mm) and curve with radius $R$ (from 40 mm to 50 mm), shown as Fig. 5b. It is very easy to observe that even though our device has relative large errors on the geometry structure parameters ($\delta$ and $R$), intensity of SPPs of output SPPs waveguide still relatively maintain at good performance. Therefore, our adiabatic device is also robust device against geometry structure parameters and fabrication of our device do not require high precision processing of manufacture to achieve low-cost device and high fidelity device.

At the last of this section, we propose a possible fabrication processing to manufacture our designed device. The fabrication techniques of multi-layered or 3D metamaterials have been widely reported. The two key issues in fabricating the coupler we designed are i) how to transfer of metallic corrugated patterns and ii) how to stack the curved waveguides. The optional methods for the former process are prevalent shadow mask lithography, soft lithography or nanoimprint lithography \cite{Walia2015,Moser2012}, and all techniques can guarantee the high-resolution in sub-microscale. As for the assembly, we can choose the suitable flexible polymer as the substrate, and peel off the structure after every metallic layer is transferred. Since our coupler can work in wide band, its outstanding superiority is the high tolerance for the structural imperfection induced by the fabrication.

\section{Conclusion}
Based on stimulated raman adiabatic passage (STIRAP) quantum control technique, we have proposed a novel coupler using a tri-layered surface plasmon-polaritons (SPPs) waveguide curved configuration, in which SPPs can be completely transferred from input corrugated thin film to output corrugated thin film in terahertz (THz) region.  We demonstrate that our design realizes highly efficient transfer with strong robustness against the perturbations of geometry parameters, and also illustrate that our device has good performance at broadband excited THz waves. This finding will make contribute to develop compact and robust integrated THz devices, which will promote the future applications in all-optical network and THz communications.

\section*{Acknowledgements}
This work is acknowledged for funding National Science and Technology Major Project (grant no. 2017ZX02101007-003); National Natural Science Foundation of China (grant no. 61565004); National Natural Science Foundation of China (grant no. 61665001); Natural Science Foundation of Guangxi Province (Nos. 2017GXNSFBA198116, 2018GXNSFAA281163).

\end{document}